\documentclass[floatfix,showpacs,superscriptaddress,prl]{revtex4-2}
\usepackage{amsbsy,amssymb,amsmath,bm}
\usepackage{float}
\usepackage{epsf}
\usepackage{graphicx,color}
\usepackage[caption = false]{subfig}
\usepackage{amsfonts}
\usepackage{amssymb}
\usepackage{amsmath}
\usepackage[T1]{fontenc}
\usepackage{xspace}
\usepackage{epsfig}
\usepackage{ctable}
\usepackage{enumitem}
\newcommand{\rrangle}{>\kern-1.2ex~>\xspace}

\definecolor{red}{rgb}{1,0,0}           
\definecolor{green}{rgb}{0,1,0}
\definecolor{blue}{rgb}{0,0,1}
\definecolor{darkblue}{rgb}{0,0,0.5}
\definecolor{lightblue}{rgb}{.5,.5,1}
\definecolor{lightgray}{gray}{.87}          
\definecolor{Dark}{gray}{.20}
\definecolor{pink}{rgb}{.95,0.82,0.92}  
\definecolor{yellow}{rgb}{1,1,0}
\definecolor{lightyellow}{rgb}{1,1,.5}
\definecolor{purple}{rgb}{0.7,0,0.85}
\definecolor{darkgreen}{rgb}{0,0.5,0}
\definecolor{orange}{rgb}{0.8,0.2,0.2}

\def \be {\bea}
\def \ee {\eea}
\def \bea {\begin{eqnarray}}
\def \eea {\end{eqnarray}}
\def \bse {\begin{subequations}}
\def \ese {\end{subequations}}
\def \bde {\begin{description}}
\def \ede {\end{description}}
\def \bee {\begin{enumerate}}
\def \eee {\end{enumerate}}
\def \nn {\nonumber}

\def \la {\langle}
\def \ra {\rangle}

\def \pd {\partial}

\def \Gt {\tilde{G}}
\def \Ht {\tilde{H}}

\def \tt {\tilde{t}}

\def \a {\alpha}
\def \b {\beta}
\def \ta {\tilde{\alpha}}
\def \tb {\tilde{\beta}}
\def \tg {\tilde{\gamma}}
\def \bg {\bar{\gamma}}
\def \td {\tilde{\delta}}
\def \dag {\dagger}
\def \g {\gamma}

\def \d {\delta}

\def \eps {\epsilon}
\def \et {\eta}

\def \n {\nu}

\def \om {\omega}
\def \Om {\Omega}

\def \th {\theta}

\def \tA {\tilde{A}}
\def \tB {\tilde{B}}
\def \tC {\tilde{C}}
\def \tD {\tilde{D}}

\def \tg {\tilde{\gamma}}

\def \HH {\mathcal{H}}

\begin{document}

\title{Winding number and Zak phase in multi-band SSH models}

\author{Chen-Shen Lee}
\affiliation{\textit{Physics Department, National Taiwan Normal University, Taipei 11677, Taiwan}}
\author{Iao-Fai Io}
\affiliation{\textit{Physics Department, National Taiwan Normal University, Taipei 11677, Taiwan}}
\author{Hsien-chung Kao\footnote{e-mail address: hckao@ntnu.edu.tw}}
\affiliation{\textit{Physics Department, National Taiwan Normal University, Taipei 11677, Taiwan}}

\date{\today }

\begin{abstract}

We use multi-band SSH models to demonstrate a prescription for calculating the correct Zak phase and winding number of multi-band systems. We verify our prescription by comparing the resultant winding number of a four-band SSH model with the edge states of a semi-infinite chain which we find by solving the equation of motion. We then also carry out an extensive comparison with the numerical results obtained by solving the matrix eigenvalue problem of finite chains with various number of sites. As a double check of our prescription, we also confirm the bulk edge correspondence in a six-band SSH model. Similar to the usual SSH model, the winding numbers associated to the left and right boundaries in a finite chain may be different if space inversion symmetry is violated in the system.  We believe the prescription we propose here may also be applied to other 1D multi-band systems.
\end{abstract}

\pacs{73.20.At,74.25.F-,73.63.Fg}
\maketitle

\subsection{I. Introduction}

Bulk-edge correspondence is one of the most notable signature of topological materials, which have been intensively studied by physicists since they were discovered \cite{Review}. The SSH model describes a one-dimension (1D) chain of polyacetylene, on which electrons hop with staggered hopping amplitudes \cite{SSH}. The system has the 1D winding number $\n$ as its corresponding topological invariant, and is arguably the simplest system with a topological phase.  Hence, it has often been used as a toy model to illustrate the properties of this kind of systems.  In particular, the number of edge states on a boundary of the system has a one-to-one correspondence to the winding number associated to the boundary see Ref.~\cite{Asboth} and references therein.

There are two sites in each unit cell of the SSH model, and thus it is a two-band model. However, it was pointed out that polyacetylene may also form a stable  trimerized chain \cite{Trimer}. To describe such a system, it is natural to consider generalizing the SSH model to a three-band one, which has been studied in Ref.~\cite{Three-band}. There is no chiral symmetry in a generic three band SSH model, and the winding number contributed by a single band is thus generally not an integer. For such a system to have a well-defined topological invariant, it is believed that one has to focus on the case that there is a space inversion symmetry in the system. Moreover, ladder-like structures of polyacetylene have also been investigated in the literature as well due to its excellent optoelectronic properties \cite{PA-ladder}. This would of course necessitate more study of multi-band topological systems, which has been relatively scare so far.

It is known that 1D winding number in the SSH model is closely related to the Zak phase \cite{Zak}, which is basically similar to the Berry phase \cite{Berry}. It has been measured directly by using a 1D optical lattice to simulate the SSH model \cite{Atala}.  By adding an on-site energy term in the SSH model, one may extend it to the Rice-Mele model \cite{Rice-Mele}. It has been shown that the Zak phase in this model is no longer quantized, since the chiral and space inversion symmetries are both violated \cite{Atala, Rhim}. One may then relate the SSH model to a Chern insulator by considering a charge-pumping process in such a system \cite{Thouless, polarization}. In fact, a relation between the 1D winding number and 2D Chern number may be established, if the 2D system has certain space inversion symmetry \cite{Ext-SSH}. Making use of such a relation, one may achieve some understanding of a 2D topological system by looking into the associated 1D system, which is much simpler in general.

It has been shown in the literature that there is an ambiguity in defining the Zak phase, $\g$, due to its dependence on the
real-space origin and unit cell \cite{Kudin, Rhim}.  It is proposed in Ref.~\cite{Kudin} that one may decompose the Zak phase into the {\it intracellular} and {\it intercellular} ones. The ambiguity only affect the intracellular Zak phase. In contrast, the intercellular Zak phase is well-defined and the bulk edge correspondence is given by $n_s = \g^{\rm inter}/\pi\; ({\rm mod}\; 2)$, with $n_s$ the number of edge states \cite{Rhim}.  Although this seems to resolve the ambiguity issue in a two band system, such a difficulty remains for generic multi-band systems, since the contribution to  $\g^{\rm inter}$ from a band may be as large as $2\pi$. As there are usually more than one occupied bands in this kind of topological systems, holonomies or equivalently a Wilson loop must be used to describe their topological structures \cite{Holonomy, W-loop}. This is effectively the non-Abelian generalization of the Berry phase \cite{Zak, Berry}.  It certainly has a much richer structure than its Abelian counterpart and has a wide application in physics \cite{H-application}.  Therefore, it is desirable to gain a deeper understanding about the non-Abelian Berry phase using a simple system such as a multi-band SSH model.

Because of the simplicity of the SSH model, we think it is worthwhile to study multi-band SSH models so that we may come up with a ``recipe'' which leads to the correct Zak phase and winding number. We believe that the lessons we learn from such a study may serve as a stepping stone for other 1D and 2D multi-band topological systems in the future. The rest of the paper is organized in the following way. In Sec. II, we focus on the four-band SSH model. We first carry out a detailed analysis of the edge states of a semi-infinite chain. We then propose a ``recipe'' to resolve the gauge ambiguity. We consider the half-filling case throughout the paper and calculate the winding number of the system analytically by summing up the contribution from the lower half of bands. By making use of the secular equation ingeniously, we are able to carry out the calculation without having to solve the secular equation explicitly. We confirm the correctness of our proposal by comparing the winding number with the number of edge states in a semi-infinite chain. Similar to the usual SSH model, there are both the left and right boundaries in a finite chain. For the bulk-edge correspondence to work properly, we must choose the four sites by the left and right edges to form the unit cells accordingly.  In Sec. III, we carry out numerical calculations by solving the matrix eigenvalue problem of a finite chain. To confirm the validity of our prescription, we check the bulk-edge correspondence in the system by comparing the two winding numbers associated to the left and right boundaries to the number of edge states seen in the numerical studies. By varying the total number of sites, we go through a comprehensive comparison of the analytical and numerical results. This provides a very strict check on the recipe that we propose. As a double check, we also use the same method to carry out similar analysis for the six-band SSH model in Appendix B. Finally, we make conclusion and discuss possible topics for future studies in Sec. IV.

\subsection{II. Edge states and winding numbers of the four-band SSH models}

Let's begin with a right semi-infinite chain of four-band SSH models starting from the site $A_1$:
\bea
&\;& \hskip -3.1cm H_{\rm SSH-4}=\sum_{j=1}^{\infty}\left\{ t_0  {\bf A}_j^\dag {\bf B}_j  +t_1 {\bf B}_j^\dag {\bf C}_j + t_2 {\bf C}_j^\dag {\bf D}_j + t_3 {\bf D}_j ^\dag  {\bf A}_{j+1}  \right\} + \mbox{ h. c.}.
\eea
Here, $j$ denotes the unit cell.  $t_0, t_1, t_2$ are the intra-cell hopping amplitudes and $t_3$ the inter-cell hopping amplitude, respectively.  This is a generalization from the dimer and trimer models for polyacetylene chains and it is a four-band model. From our experience with the SSH model, it is expected that the 1D winding number may now range from $0$ to $3$.

The corresponding energy eigenstates would satisfy the following equation of motion(EOM):
\bea \label{SSH4 EOM 1}
&\;& \hskip -3.1cm \Om A_j - \left(t_0 B_j + t_3 D_{j-1} \right) =0, \cr
&\;& \hskip -3.1cm \Om B_j - \left(t_0 A_j + t_1 C_j \right) =0, \cr
&\;& \hskip -3.1cm \Om C_j - \left(t_1 B_j + t_2 D_j \right) =0, \cr
&\;& \hskip -3.1cm \Om D_j - \left(t_2 C_j + t_3 A_{j+1}\right) =0,
\eea
where $\Om$ denotes the energy of the system.  The EOM is a linear recurrence relation with constant coefficients.  As the $A_1$ site is the leftmost one, it is impossible for electrons to hop further to the left on this site.  Hence, the open boundary condition (OBC) is given by
\bea \label{SSH4 BC 1}
&\;& \hskip -3.1cm \Om A_1 - t_0 B_1 = 0, \nn
\eea
which may be simplified to
\bea \label{SSH4 Simplified BC 1}
&\;& \hskip -3.1cm D_0 =0.
\eea
To find the solutions to the EOM, we let
\bea
&\;& \hskip -3.1cm A_j =\a s^j, B_j=\b s^j, C_j =\g s^j, D_j=\d s^j.
\eea
As a result, we have
\bea \label{SSH4 abcd 1}
&\;& \hskip -3.1cm \Om \a - \left( t_0 \b + t_3 s^{-1} \d \right) = 0, \cr
&\;& \hskip -3.1cm \Om \b - \left( t_0 \a + t_1 \g \right) = 0, \cr
&\;& \hskip -3.1cm \Om \g - \left( t_1 \b + t_2 \d \right) = 0, \cr
&\;& \hskip -3.1cm \Om \d - \left( t_2 \g + t_3 s \a \right)= 0.
\eea
We may have non-trivial solutions for $\a, \b, \g$ and $\d$ only if the determinant formed by the coefficients of them is vanishing, i.e.
\bea \label{SSH4 Secular eq 1}
&\;& \hskip -3.1cm \Om^4 - \left(t_0^2 + t_1^2 + t_2^2 + t_3^2 \right)\Om^2 + \left( t_0^2 t_2^2 + t_1^2 t_3^2 - 2 t_0 t_1 t_2 t_3 u \right) = 0,
\eea
with $u = \left( s + s^{-1} \right)/2.$  We will refer to this as the secular equation in this article. Note that the equation is quartic in $\Om$, so there are four roots of $\Om$ to the equation. Since the above equation is quadratic in $s$ and is symmetric with respect to space inversion, $s \to s^{-1}$, the most general forms of $A_j, B_j, C_j$ and $D_j$ are given by
\bea \label{SSH4 general solution}
&\;& \hskip -3.1cm A_j = \a_+ s^j + \a_- s^{-j}, B_j = \b_+ s^j + \b_- s^{-j},  C_j = \g_+ s^j + \g_- s^{-j},  D_j = \d_+ s^j + \d_- s^{-j}.
\eea
If we assume $|s|<1$ and focus on the edge states, then physical solutions exist only if $\a_- = \b_- = \g_- = \d_- =0$. Combining this with the boundary condition (BC) $D_0 =0$, we see in turn that $\d_+ = 0$. As a result, eq.~(\ref{SSH4 abcd 1}) reduces to
\bea \label{SSH4 abcd edge}
&\;& \hskip -3.1cm \Om \a_+ - t_0 \b_+  = 0, \cr
&\;& \hskip -3.1cm \Om \b_+ - \left( t_0 \a_+ + t_1 \g_+ \right) = 0, \cr
&\;& \hskip -3.1cm \Om \g_+ -  t_1 \b_+ = 0, \cr
&\;& \hskip -3.1cm  t_2 \g_+ + t_3 s \a_+ = 0.
\eea
There are four equations with only three independent variables $\a_+, \b_+$ and $\g_+$. By choosing arbitrary three equations among the four, we may obtain four $3\times 3$ determinants formed by these coefficients. To have non-trivial solutions, all the four determinants must be vanishing.  By counting the degrees of freedom, one see that only two among the four conditions are independent.  For simplicity, let's choose the first three equations to achieve the determinant for the first condition, and choose the first, the second , and the last equations to achieve that for the second condition.  They take the following forms:
\bea \label{SSH4 Edge state}
&\;& \hskip -3.1cm \Om \left( \Om^2 - t_0^2 - t_1^2 \right)  = 0, \cr
&\;& \hskip -3.1cm  t_0 \left( t_0 t_2 - t_1 t_3 s \right) - t_2 \Om^2 = 0.
\eea
The solutions to the above equations are given by
\bea \label{SSH4 Edge state solution}
&\;& \hskip -3.1cm s = \frac{t_0 t_2}{t_1 t_3},\;  \Om = 0, \cr
&\;& \hskip -3.1cm s = -\frac{t_1 t_2}{t_0 t_3},\;  \Om =\pm \sqrt{ t_0^2 + t_1^2}.
\eea
From the above results, we see there are indeed four phases:
\bee[label=\arabic*)]
\item $\left| \frac{t_0 t_2}{t_1 t_3} \right|<1, \left| \frac{t_1 t_2}{t_0 t_3} \right|<1$, three edge states, one of which is a zero energy state.

\item $\left| \frac{t_0 t_2}{t_1 t_3} \right|>1, \left| \frac{t_1 t_2}{t_0 t_3} \right|<1$, two edge states.

\item $\left| \frac{t_0 t_2}{t_1 t_3} \right|<1, \left| \frac{t_1 t_2}{t_0 t_3} \right|>1$, one zero energy edge state.

\item $\left| \frac{t_0 t_2}{t_1 t_3} \right|>1, \left| \frac{t_1 t_2}{t_0 t_3} \right|>1$, no edge state.
\eee
From the bulk edge correspondence, we expect the winding numbers $\n$ of the system to be $3, 2, 1$ and $0$, respectively.  To calculate $\n$ or equivalently the Zak phase \cite{Zak}, which may be used to classify the system, we have to impose the periodic boundary condition. Consequently, we obtain the following Bloch Hamiltonian:
\be\label{SSH-4}
\HH_{\rm SSH-4}=\left(
\begin{matrix}
0 & t_0 & 0 & t_3 {\rm e}^{-ip} \cr
t_0 & 0 & t_1 & 0 \cr
0 & t_1 & 0 & t_2  \cr
t_3 {\rm e}^{ip} & 0 & t_2 & 0 \cr
\end{matrix}
\right),
\ee
where $p$ is the lattice momentum.  The chiral operator that anti-commutes with $\HH_{\rm SSH-4}$ is given by
\be\label{Pi}
\Pi=\left(
\begin{matrix}
1 &  0 & 0 &  0 \cr
0 & -1 & 0 &  0 \cr
0 & 0  & 1 &  0 \cr
0 &  0 & 0 & -1 \cr
\end{matrix}
\right).
\ee
Its existence grantees that eigenstates of $\HH_{\rm SSH-4}$ with non-zero energy would always appear in pairs with eigenvalues $(\Om, -\Om)$ and the corresponding eigenstates are related by $\left|-\Om \right\ra = \Pi \left|\Om \right\ra$. In contrast, zero energy eigenstates can always be chosen to be chiral eigenstates and the left-handed and right-handed states are decoupled from each other.  Thus, this kind of systems belong to the BDI class \cite{Periodic-table}.

In this case, eqs.~(\ref{SSH4 Secular eq 1}) becomes
\bea \label{SSH4 Secular eq 2}
&\;& \hskip -3.1cm \Om^4(p) - \left(t_0^2 + t_1^2 + t_2^2 + t_3^2 \right)\Om^2(p) + \left( t_0^2 t_2^2 + t_1^2 t_3^2 - 2 t_0 t_1 t_2 t_3  \cos p \right) = 0
\eea
The four solutions are given by $\Om_1(p) = -\om_2(p), \Om_2(p) = -\om_1(p), \Om_3(p) = \om_1(p)$, and $\Om_4(p) = \om_2(p)$, with
\bea
\om_{1,2}(p)= \Biggl\{ \frac{ \left(t_0^2 + t_1^2 + t_2^2 + t_3^2 \right) \mp \Bigl[  \left(t_0^2 + t_1^2 + t_2^2 + t_3^2 \right)^2 - 4 \left( t_0^2 t_2^2 + t_1^2 t_3^2 - 2 t_0 t_1 t_2 t_3  \cos p \right) \Bigr]^{1/2} }{2}  \Biggr\}^{1/2}.
\eea
Since we focus on the half-filled case, we must sum over the lower two bands to achieve the topological invariant $\n$ here in contrast to the case of two-band models. The analytical expression for the 1D winding number $\n$ is given by
\bea\label{1d-winding-number}
\n = \sum_{a =1,2} \frac{i}{2\pi} \int_{-\pi}^{\pi} dp\, \Bigl\{ \left\la \Om_a (p) \right | \left. \pd_p \Om_a(p) \right\ra - \left\la \pd_p \Om_a (p) \right | \left. \Om_a(p) \right\ra \Bigr\},
\eea
where $\left| \pd_p \Om_a(p) \right\ra \equiv \pd_p  \left| \Om_a(p) \right\ra$, with $a$ the band index \cite{Zak, Berry, TKNN}.

To proceed, we must find the corresponding wave functions:
\bea \label{SSH4 abcd 2}
&\;& \hskip -3.1cm \left|\Om_a(p) \right\ra ={\cal N}_a(p) \left\{ \ta_a(p), \tb_a(p), \tg_a(p), \td_a(p) \right\}^T.
\eea
$\ta_a(p), \tb_a(p), \tg_a(p),$ and $\td_a(p)$ satisfy eq.~(\ref{SSH4 abcd 2}) and $a=1, \ldots, 4$. It is known in the literature that there exist ambiguity in determining the Zak phase, $\g$, in 1D topological systems \cite{Atala, Kudin, Rhim}. It is proposed that such an ambiguity may be resolved by separating it into the intracellular and intercellular Zak phases. It has been shown that the intercellular Zak phase is origin-independent. Hence, it is suggested that the bulk edge correspondence is given by $n_s = \g^{\rm inter}/\pi\; ({\rm mod}\; 2)$, with $n_s$ the number of edge states. This, however, cannot be applied to a generic multi-band system, since the contribution to  $\g^{\rm inter}$ from a band may be as large as $2\pi$.

Let's now spell out the details of our ``recipe'' to achieve the correct winding number. From eq.~(\ref{SSH4 abcd 2}), we see that $\ta_a(p), \tb_a(p), \tg_a(p),$ and $\td_a(p)$ may be expressed as
\bea
&\;& \hskip -3.1cm \ta_a(p) = \Bigl\{ t_3 \left( \Om_a^2(p) - t_1^2 \right) {\rm e}^{-ip} + t_0 t_1 t_2  \Bigr\}, \cr
&\;& \hskip -3.1cm \tb_a(p) = \Om_a(p) \Bigl\{  - t_1 t_2 - t_0 t_3 {\rm e}^{-ip}  \Bigr\} , \cr
&\;& \hskip -3.1cm \tg_a(p) = \Bigl\{ t_2 \left( \Om_a^2(p) - t_0^2 \right) + t_0 t_1 t_3 {\rm e}^{-ip}  \Bigr\}, \cr
&\;& \hskip -3.1cm \td_a(p) =  \Om_a(p)\Bigl\{ t_0^2 + t_1^2 - \Om_a^2(p) \Bigr\}.
\eea
Here, ${\cal N}_a(p)$ is a proper normalization constant that enforces the condition $\left\la \Om_a(p)\, \right|\left. \Om_a(p) \right\ra = 1.$  In other words,
\bea
&\;& \hskip -3.1cm {\cal N}_a(p) = \left\{ \Bigl| \ta_a(p) \Bigr|^2 + \left|\tb_a(p) \right|^2 + \Bigl|\tg_a(p) \Bigr|^2 + \left|\td_a(p) \right|^2 \right\}^{-1/2}.
\eea
Here, we have deliberately chosen $\td_a(p)$ to be real, which is equivalent to choosing a gauge. The reason for such a choice is that we have in mind to express $\ta_a(p), \tb_a(p)$ and $\tg_a(p)$ in terms of $\td_a(p)$, which is in turn dictated by the BC on the left edge: $D_0 = 0$.  To confirm that this is indeed the correct gauge choice, we will compare the resultant winding number to the number of edge states both analytically and numerically later on.  We would like to emphasize here that although the bulk edge correspondence is expected to hold in a topological system, it is by no means trivial to come up with a viable prescription for the correct form of the winding number so that such a correspondence is enforced explicitly in a multi-band system.

We relegate the details of the calculation to Appendix A and would just give the final expression for the winding number here.
\bea\label{Winding-number-SSH4}
&\;& \hskip -3.1cm \n = \frac{1}{\pi} \int_{-\pi}^{\pi} dp\, \Biggl\{ \frac{ t_0 t_3 {\rm e}^{ip} }{ t_1 t_2 + t_0 t_3 {\rm e}^{ip} } + \frac{ t_1 t_3 {\rm e}^{ip} }{ 2\Bigl( t_0 t_2 + t_1 t_3 {\rm e}^{ip}\Bigr) } \Biggr\}, \cr
&\;& \hskip -3.1cm  = 2\th \left( |t_0 t_3| - |t_1 t_2| \right) + \th \left( |t_1 t_3| - |t_0 t_2| \right),
\eea
where $\th(x)$ is the step function. One can easily check that the above result is identical to that obtained by analyzing the edge states.

To verify the above results numerically, we must consider a finite chain of four-band SSH model. Let's start with a system consisting of $N$ unit-cells, i. e. $4N$ sites:
\bea
&\;& \hskip -3.1cm H_{\rm SSH}=\sum_{j=1}^{N-1}\left\{ t_0  {\bf A}_j^\dag {\bf B}_j  +t_1 {\bf B}_j^\dag {\bf C}_j + t_2 {\bf C}_j^\dag {\bf D}_j + t_3 {\bf D}_j ^\dag  {\bf A}_{j+1}  \right\} + t_0  {\bf A}_N^\dag {\bf B}_N  +t_1 {\bf B}_N^\dag {\bf C}_N + t_2 {\bf C}_N^\dag {\bf D}_N + \mbox{h. c.}.
\eea
It is straightforward to see that the energy eigenstates would satisfy the following EOM:
\bea \label{SSH4 EOM 4}
&\;& \hskip -3.1cm \Om A_j - \left(t_0 B_j + t_3 D_{j-1} \right) =0, \quad \mbox{\rm for }  j = 2, \ldots,N; \cr
&\;& \hskip -3.1cm \Om B_j - \left(t_0 A_j + t_1 C_j \right) =0, \quad\quad \mbox{\rm for }  j = 2, \ldots,N; \cr
&\;& \hskip -3.1cm \Om C_j - \left(t_1 B_j + t_2 D_j \right) =0, \quad\quad \mbox{\rm for }  j = 2, \ldots,N; \cr
&\;& \hskip -3.1cm \Om D_j - \left(t_2 C_j + t_3 A_{j+1}\right) =0, \quad \mbox{\rm for }  j = 1, \ldots, N-1.
\eea
Meanwhile, the BC's are
\bea \label{SSH4 Simplified BC 4}
&\;& \hskip -3.1cm D_0 =0; \cr
&\;& \hskip -3.1cm A_{N+1} = 0.
\eea
By using the most general forms of $A_j, B_j, C_j$ and $D_j$ given in eq.~(\ref{SSH4 general solution}), the above BC's become
\bea
&\;& \hskip -3.1cm \d_+ + \d_- = 0; \cr
&\;& \hskip -3.1cm \a_+ + \a_- = 0.
\eea
From eq.~(\ref{SSH4 abcd 2}), we may  express $\a_I$ in terms of $\d_I$:
\bea \label{SSH4 abcd 3}
&\;& \hskip -3.1cm \a_I = \frac{ \left( t_3 \Om^2 - t_1^2 t_3 + t_0 t_1 t_2 s_I \right) \d_I}{s_I \Om\left( t_0^2 + t_1^2 - \Om^2 \right)},
\eea
where $I = +, -$ and $s_+ = s, s_- = s^{-1}.$ Substitute the above results to the BC's and they may then be simplified to
\bea
&\;& \hskip -3.1cm \d_+  + \d_-  = 0; \cr
&\;& \hskip -3.1cm \left( t_3 \Om^2 - t_1^2 t_3 + t_0 t_1 t_2 s \right)s^{N} \d_+ + \left( t_3 \Om^2 - t_1^2 t_3 + t_0 t_1 t_2 s^{-1} \right)s^{-N} \d_- = 0.
\eea
Non-trivial solutions for $\d_+$ and $\d_-$ exist only if the determinant formed by them is vanishing. This leads to the characteristic equation of $s$:
\bea
&\;& \hskip -3.1cm t_3 \left( \Om^2 - t_1^2 \right)U_{N-1}(u) + t_0 t_1 t_2 U_{N}(u)  = 0,
\eea
where $U_N(u)$ is the Chebyshev polynomials of the second kind of order $N$.
Combining this with the secular equation in (\ref{SSH4 Secular eq 1}), we may find $u$ and the corresponding energy. Of course, this can only be done numerically.  We may compare the energy spectrum and wave functions obtained in this way with those obtained by numerical diagonalization of the Hamiltonian matrix. Hence, it provides a double check on the numerical results.

Since there is also a right boundary in a finite chain, there would be right edge states as well.  To find these associated edge states and their energies, we must consider a left semi-infinite chain of four-band SSH models with OBC ending at the site $D_N$. Upon space inversion, we now rename $D_N, C_N, B_N, A_N$ to be $\tA_1, \tB_1, \tC_1, \tD_1$. As a result,  we may achieve the right edge states from the results of the right semi-infinite chain easily by identifying $\tt_0 = t_2, \tt_1 = t_1, \tt_2 = t_0$ and $\tt_3 = t_3$. Here, $\tt_0, \tt_1, \tt_2, \tt_3$ are the hopping amplitudes between $(\tA, \tB), (\tB, \tC), (\tC, \tD)$ and $(\tD, \tA)$ sites, respectively. Consequently, we have for the right boundary
\bee[label=\arabic*)]
\item $\left| \frac{t_0 t_2}{t_1 t_3} \right|<1, \left| \frac{t_1 t_0}{t_2 t_3} \right|<1$, three edge states, one of which is a zero energy state.

\item $\left| \frac{t_0 t_2}{t_1 t_3} \right|>1, \left| \frac{t_1 t_0}{t_2 t_3} \right|<1$, two edge states.

\item $\left| \frac{t_0 t_2}{t_1 t_3} \right|<1, \left| \frac{t_1 t_0}{t_2 t_3} \right|>1$, one zero energy edge state.

\item $\left| \frac{t_0 t_2}{t_1 t_3} \right|>1, \left| \frac{t_1 t_0}{t_2 t_3} \right|>1$, no edge state.
\eee
If $t_0 = t_2$ then the system is inversion symmetric with respect to the center of a unit cell. When this is the case, the numbers of edge states on the left and right boundaries would be the same. Moreover, the Zak phase or equivalently the winding number contributed by each band would be quantized, as predicted in Ref.~\cite{Space-inversion}.  In contrast, if $t_0 \neq t_2$ such a symmetry would be spoiled and the numbers of edge states on the left and right boundaries would generally be different. Thus, it is possible that the two boundaries of the system are in different phases. Furthermore, the Zak phase contributed by each band would not be quantized in general. This is very different from the two-band SSH model, and may only appear in a system with four or more bands. The conclusion that we achieve here may be used to understand some of the results obtained in Ref.~\cite{Trimer}.  Applying the analysis for the edge states that we carried out previously to the three band SSH model, it may be shown that $\n_{\rm L} = 2 \th \left( t_2-t_1 \right)$ and $\n_{\rm R} = 2 \th \left( t_0-t_1 \right)$ for the case that $N_{\rm tot} = 3N$.  As we identify the parameters $(t_0, t_1, t_2)$ with $(u, v, w)$ in the article, we see immediately why there are three phases when we vary $w$ while setting $u=1$ and $v=2$.

We would like to point out this kind of phenomenon can only appear multi-band SSH models. The reason is that the usual SSH has both chiral and space inversion symmetries, and it is impossible to break one of them without violating the other in such a system. On the contrary, we may separate the effect of the two symmetries in a multi-band SSH model.  In particular, a three band SSH model may only have inversion symmetry, while a four-band SSH model may only have chiral symmetry. Hence, multi-band SSH models are examples of study topological crystalline insulators \cite{TC-insulators}.

We may also consider finite chains with other total numbers of sites, $N_{\rm tot} = 4N+1, 4N+2,  4N-1$.  In these cases, even if $t_0 = t_2$ so that a complete unit cell itself is inversion symmetric, such a symmetry may still be violated generally by the BC on the right boundary.  We will just summarize the results here.
\bee[label=\arabic*)]
\item $N_{\rm tot} = 4N+1$, the right BC: $B_{N+1}=0$. $\tt_0 = t_3, \tt_1 = t_2, \tt_2 = t_1,$ and $\tt_3 = t_0$.\hfill\break
The corresponding characteristic equation is
\bea
&\;& \hskip -3.1cm t_0 t_3 U_{N-1}(u) + t_1 t_2 U_{N}(u) = 0,
\eea

\item $N_{\rm tot} =4N+2$, the right BC: $C_{N+1}=0$. $\tt_0 = t_0, \tt_1 = t_3, \tt_2 = t_2,$ and $\tt_3 = t_1$.\hfill\break
The corresponding characteristic equation is
\bea
&\;& \hskip -3.1cm   t_0 t_1 t_3 U_{N-1}(u) + t_2 \left( \Om^2 - t_0^2 \right) U_{N}(u) = 0.
\eea

\item $N_{\rm tot} = 4N-1$, the right BC: $D_{N+1}=0$. $\tt_0 = t_1, \tt_1 = t_0, \tt_2 = t_3,$ and $\tt_3 = t_2$.\hfill\break
The corresponding characteristic equation is
\bea
&\;& \hskip -3.1cm  s^N - s^{-N} = 0.
\eea
Hence,
$s= {\rm e}^{i\pi k/N}$, with $k = 1, \ldots, N-1$.
\bea
\om_{1,2}= \Biggl\{ \frac{ \left(t_0^2 + t_1^2 + t_2^2 + t_3^2 \right) \mp \Bigl[  \left(t_0^2 + t_1^2 + t_2^2 + t_3^2 \right)^2 - 4 \left\{ t_0^2 t_2^2 + t_1^2 t_3^2 - 2 t_0 t_1 t_2 t_3  \cos(\pi k/N) \right\} \Bigr]^{1/2} }{2}  \Biggr\}^{1/2}.
\eea
Here, the bulk states and their energies may be found analytically. By counting the degree of freedom, we see there are always three edge states in total. Thus, we have the relation that  $\n_{\rm R} = 3 - \n_{\rm L}$, which is a generalization of the result in the case that there are odd number of sites in a two-band SSH model.
\eee

\subsection{III. Numerical results of the four-band SSH models}

\begin{figure}[hbt!]
\centering
\includegraphics[width=0.60\textwidth]{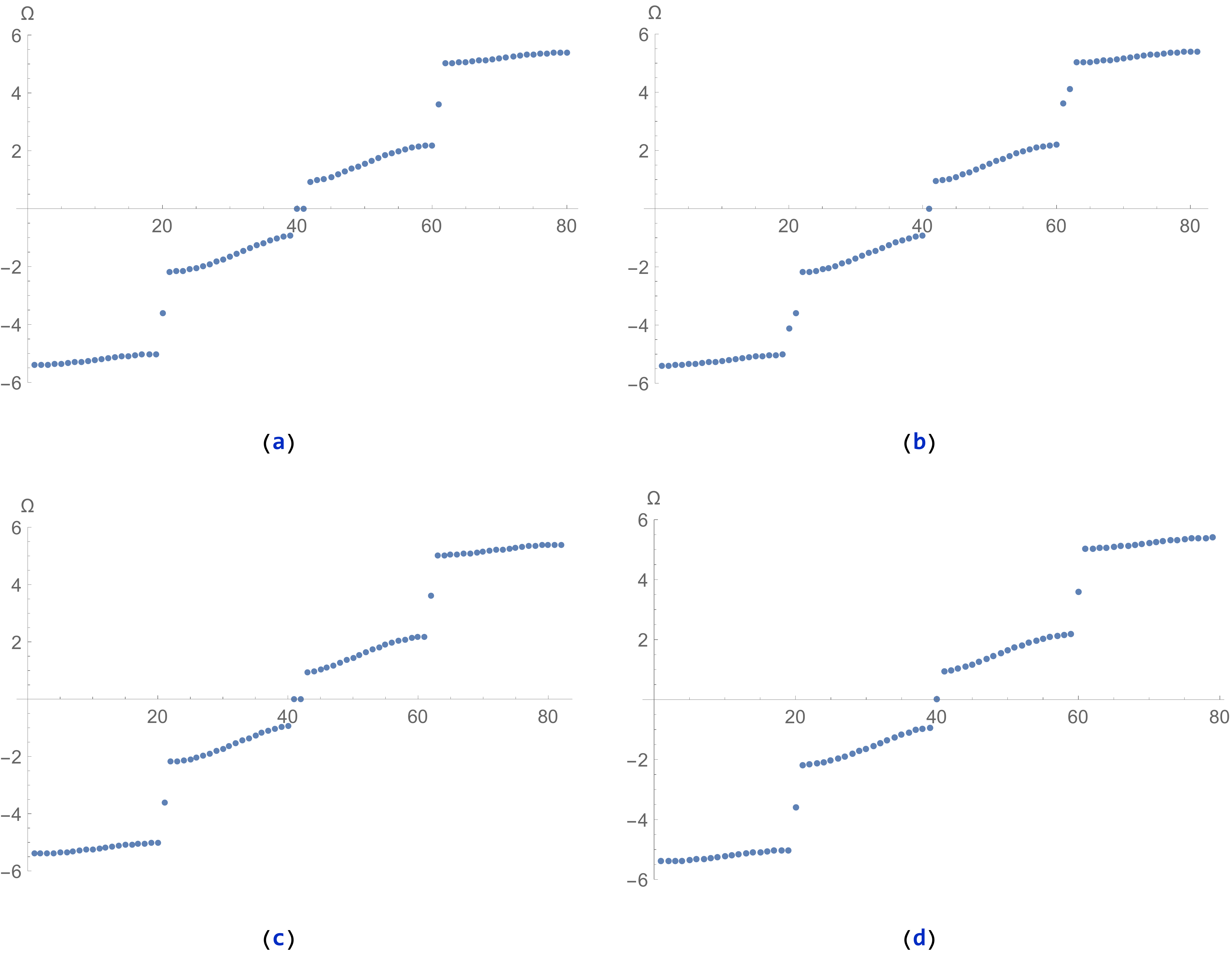}
\caption{The parameters are chosen to be $\left( t_0, t_1, t_2, t_3 \right) = (3, 2, 1, 4)$ so that  the winding numbers on the left boundary $\n_{\rm L} = 3$. The energy spectra for the four cases that $N_{\rm tot} = 80, 81, 82$ and $79$ are shown. (a) $N_{\rm tot} = 80$, $\n_{\rm R} = 1$. There are four edge states in total, two of which have approximately zero energy.  (b) $N_{\rm tot} = 81$, $\n_{\rm R} = 2$. There are five edge states in total, one of which have exact zero energy. , (c) $N_{\rm tot} = 82$, $\n_{\rm R} = 3$. There are six edge states in total, two of which have approximately zero energy.  (d)$N_{\rm tot} = 79$, $\n_{\rm R} = 0$. There are three edge states in total, one of which have exact zero energy. }  \label{fig1 SSH4}
\end{figure}

All the predictions in Sec. II may be confirmed numerically. For this purpose, the parameters are chosen to be $\left( t_0, t_1, t_2, t_3 \right) = (3, 2, 1, 4)$ so that the winding numbers on the left boundary $\n_{\rm L} = 3$. We consider the four cases that $N_{\rm tot} = 80, 81, 82$ and $79$.  The energy spectra for these cases are shown in Fig.~\ref{fig1 SSH4}.  From the criteria that we obtained previously, the winding numbers on the right boundary, $\n_{\rm R}$ should be $1, 2, 3$ and $0$, respectively. Note that even if $N_{\rm tot}$ is a multiple of four, $\n_{\rm L}$ and $\n_{\rm R}$ are different since there is no inversion symmetry in this system.  As a consequence of this, the energies of left and right edge states are generally different from each other except for the almost zero energy pair. Barring the case that $N_{\rm tot} = 82$, the bulk edge correspondence may be readily verified from the figures, since we expect the edge states to be mid-gap states. By plotting their wave functions, we confirm that there are indeed six edge states in the case that $N_{\rm tot} = 82$. Moreover, although it is a finite chain, the energies of the edge states are in good agreement with those predicted in eq.~(\ref{SSH4 Edge state solution}).  Using these parameters, we may also check the ``winding numbers'' contributed by the lower two bands are 1.60827 and 1.39173.  They are both larger than one and are not quantized. Therefore, the prescription proposed in Ref.~\cite{Rhim} cannot resolve the difficulty we encounter here.

To see the band structures clearly, we also show the dispersion relation between $\Om$ and $p$ for $N_{\rm tot} = 80, 81$. Here, the momentum $p = -i \log(s) $.  Note that the $s$'s associated to the edge states are purely real, so the corresponding momenta are either $0$ or $\pi$.
\begin{figure}[hbt!]
\centering
\includegraphics[width=0.60\textwidth]{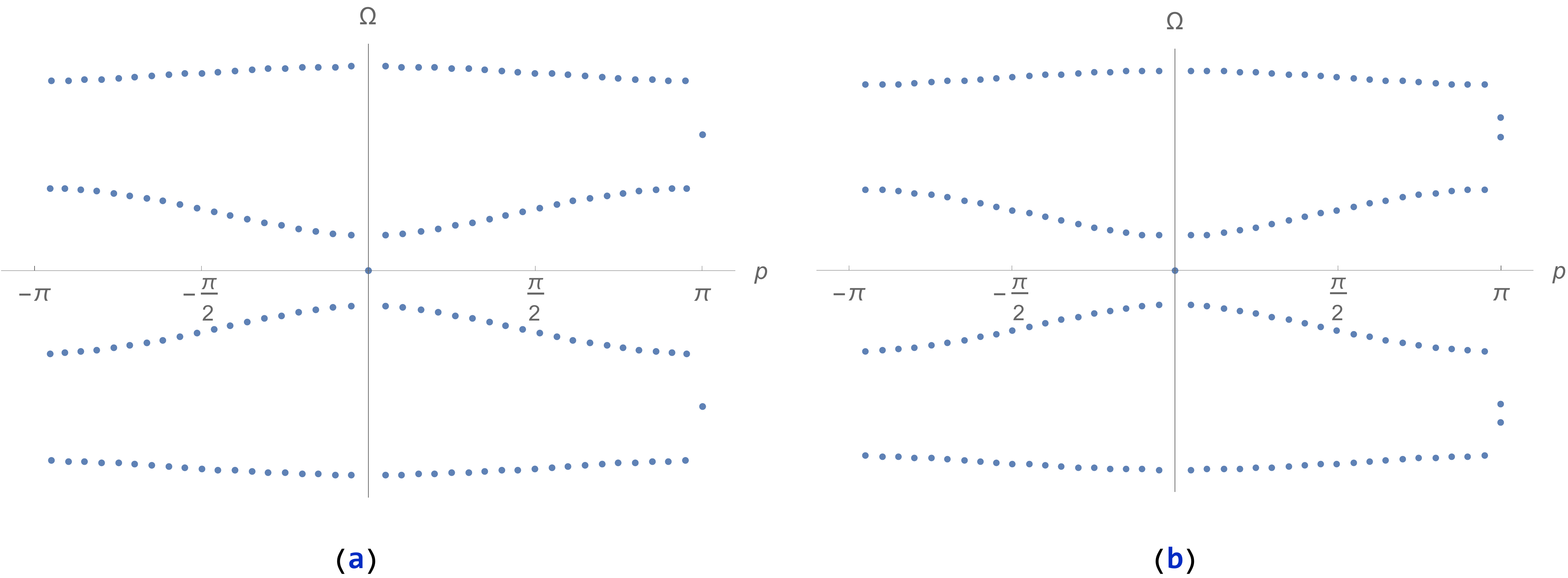}
\caption{The dispersion relation for $N_{\rm tot} = 80, 81$ are shown. (a) $N_{\rm tot} = 80$, $\n_{\rm R} = 1$.  (b) $N_{\rm tot} = 81$, $\n_{\rm R} = 2$.}  \label{fig2 SSH4}
\end{figure}

\begin{figure}[hbt!]
\centering
\includegraphics[width=0.60\textwidth]{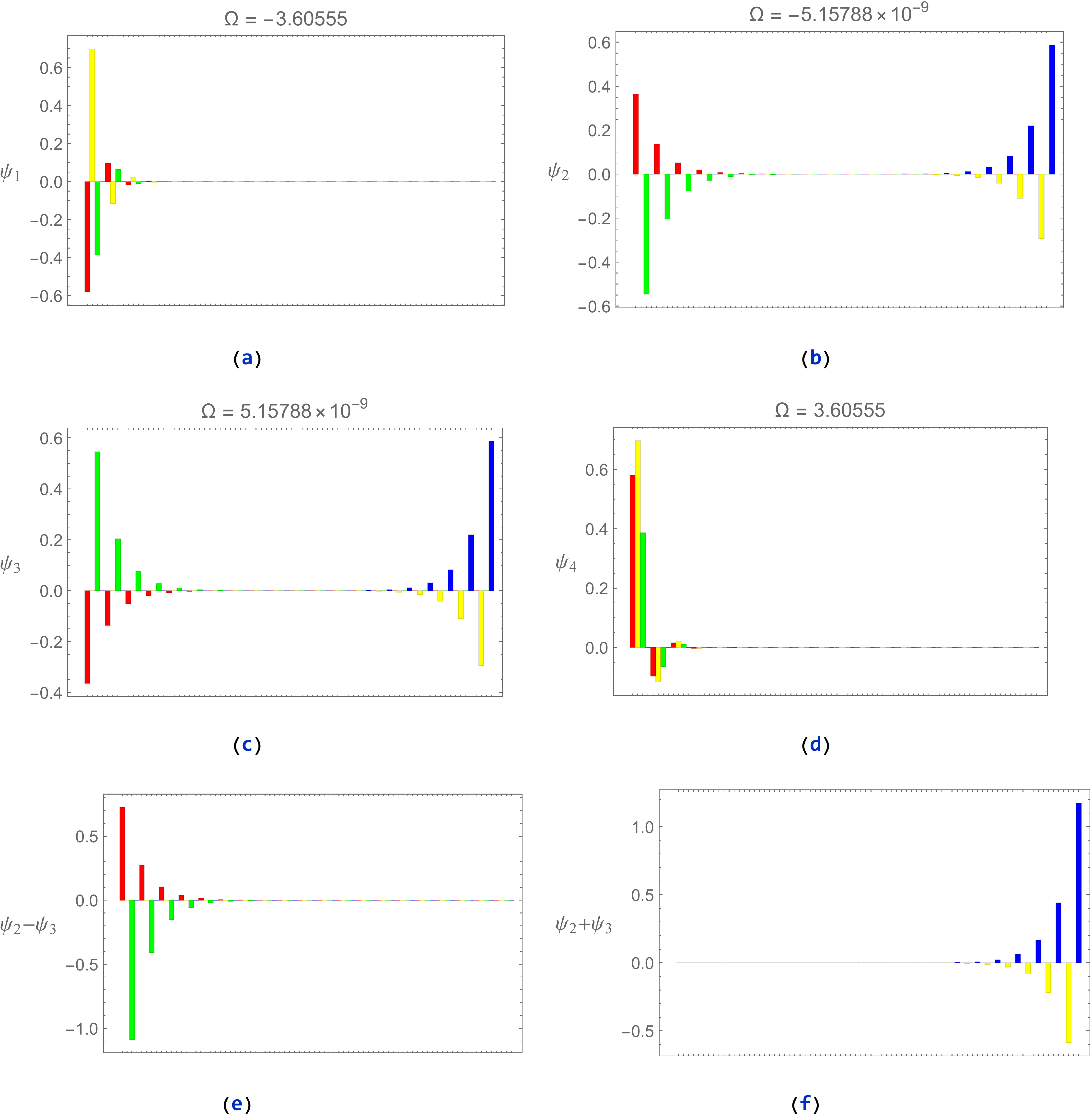}
\caption{Here we show the wave functions for the edge states in the case that $\left( t_0, t_1, t_2, t_3 \right) = (3, 2, 1, 4)$ with $N_{\rm tot} = 80$.(a) The wave functions for the edge state with the lowest energy, denoted $\psi_1$. It is a left edge state.  (b,), (c) The wave functions of the two edge states with approximately zero energy in the system, denoted $\psi_2$ and $\psi_3$. They form a chiral pair. (d)The wave functions for the edge state with the highest energy, denoted $\psi_4$. It is also a left edge state and is the chiral counterpart of $\psi_1$. (e), (f) The difference and sum of $\psi_2$ and $\psi_3$. They become purely left and right edge states, respectively.}  \label{fig3 SSH4}
\end{figure}

The wave functions for the edge states in the case that $N_{\rm tot} = 80$ are shown in Fig.~\ref{fig3 SSH4}. Here, $\n_{\rm L} = 3$ and $\n_{\rm R} = 1$. As a result, there are four edge states in total. The edge state with the lowest energy and that with the highest energy form a chiral pair.  They are both purely left edge states.  There are two approximately zero energy states, denoted $\psi_2$ and $\psi_3$, respectively. In contrast, they involve mixing of the left and right edge states due to quantum tunneling. Their difference and sum lead to left and right edge states, respectively.

\begin{figure}[hbt!]
\centering
\includegraphics[width=0.60\textwidth]{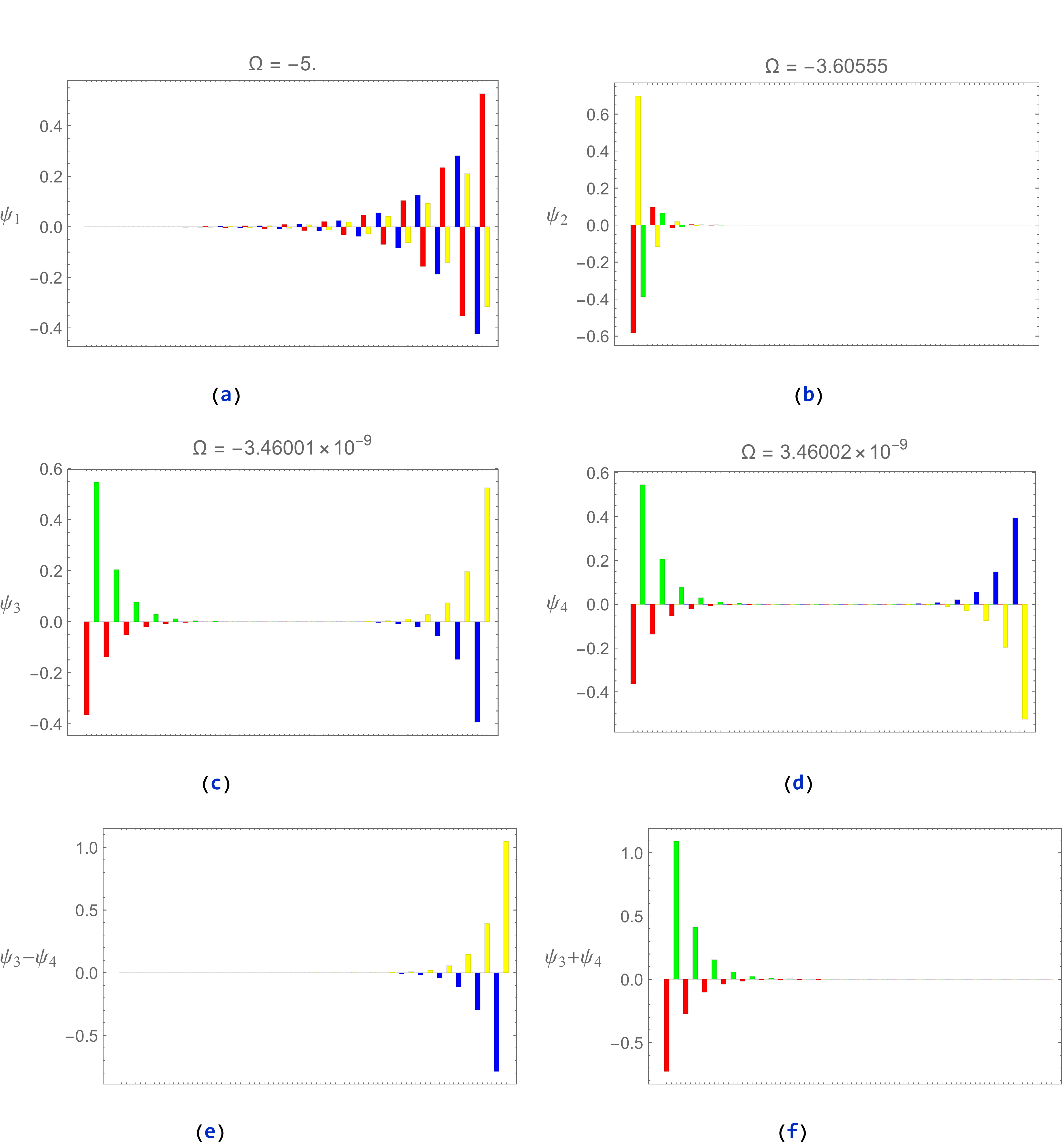}
\caption{Here we show the wave functions for the edge states in the case that $\left( t_0, t_1, t_2, t_3 \right) = (3, 2, 1, 4)$ with $N_{\rm tot} = 82$.(a) The wave functions for the edge state with the lowest energy, denoted $\psi_1$. It is a right edge state.  (b) The wave functions for the edge state with the second lowest energy, denoted $\psi_2$. It is a left edge state. (c), (d) The wave functions of the two edge states with approximately zero energy in the system, denoted $\psi_3$ and $\psi_4$. They form a chiral pair. (e), (f) The difference and sum of $\psi_3$ and $\psi_4$. They become purely right and left edge states, respectively.}  \label{fig4 SSH4}
\end{figure}

The wave functions for the edge states in the case that $N_{\rm tot} = 82$ are shown in Fig.~\ref{fig4 SSH4}. Here, $\n_{\rm L} = 3$ and $\n_{\rm R} = 3$. As a result, there are six edge states in total. To save space, we only show the wave functions of the lowest four energy edge states. The edge state with the lowest energy is a purely right edge state, while the edge state with the second lowest energy is a purely left edge state. There are two approximately zero energy states, denoted $\psi_3$ and $\psi_4$, respectively. Again, they involve mixing of the left and right edge states due to quantum tunneling and the difference and sum of them lead to right and left edge states, respectively.

Some comments are in order. First, we notice that some of the edge states in the four-band SSH model are mid-gap states with non-vanishing energy unlike those in the usual SSH model. In fact, the energies of some of the edge states may lie very close to a band edge. It is therefore not clear whether this kind of edge states are robust under perturbation. Secondly, the above results seem to suggest that
\be
N_{\rm tot} \equiv \n_{\rm L} + \n_{\rm R} \left( \mbox{mod 4} \right). \nn
\ee
However, this is nothing but a coincidence. This can be seen by choosing other sets of parameters.  In particular, we have also chosen the following three sets of parameters: $\left( t_0, t_1, t_2, t_3 \right) = (2, 3, 2, 4)$, $(3, 2, 3, 4)$, and $(3, 5, 4, 4)$ for $N_{\rm tot}=4N$. Because of the inversion symmetry, here we have $\n_{\rm L} = \n_{\rm R}=\n$, with $\n=3,2,1,$ respectively.  More specifically, the winding numbers contributed from the lower two bands are $(\n_1, \n_2) = (2, 1)$, $(1, 1)$, and $(1, 0)$. The numbers of mid-gap states lying in the three band-gaps are $(2, 2, 2)$, $(2, 0, 2)$, and $(0, 2, 0)$, respectively.  Although $\n_1$ and $\n_2$ are both quantized in this case, at this moment we cannot see any connection between them and the numbers of mid-gap states, besides the already known bulk edge correspondence.  Note that $\n_1 =2$ in the case $\left( t_0, t_1, t_2, t_3 \right) = (2, 3, 2, 4)$. This is again an explicit example that the proposal $n_s = \g^{\rm inter}/\pi\; ({\rm mod}\; 2)$ in Ref.~\cite{Rhim} does not work in a multi-band system.

Finally, as a double check of the prescription for calculating the right winding number that we put forward in this paper, we also study the six-band SSH model by following similar procedures taken previously. Since the reasoning has been laid out explicitly in the previous section, we will only summarize the main results and leave them in Appendix B. Suffice to say that the results confirm that our winding number prescription is indeed correct.

\subsection{IV. Conclusion and discussion}

In this paper, we propose a recipe to calculate the correct Zak phase and winding number for a multi-band topological system. We use the four-band SSH model to explicitly demonstrate how this may be done.  We first compare the winding numbers obtained in this way with the analysis of edge states of a semi-infinite chain of the model. To further check the validity of our prescription, we then consider a finite chain with various number of sites so that we may carry out an extensive comparison between the analytic formula and the numerical computation. We verify all the results predicted by our prescription, including the number of edge states on the left and right boundaries respectively and their corresponding energies. Thus, we are quite confident our prescription for calculating the winding numbers of multi-band models are correct. In contrast to the usual SSH model, even if $N_{\rm tot} = 4N, 6N$ in these systems, $\n_{\rm L}$ and $\n_{\rm R}$ are generally different if they do not have an inversion symmetry. Thus, we need two bulk Hamiltonians to completely describe the topological phase of a finite chain for such systems. Since little assumption has been made in our analysis of the edge states in a semi-infinite chain, the method may also be applied to investigate other 1D multi-band systems of interest.

Finally, we would like to discuss some possible directions for further study.  First, as we mentioned previously, the left and right winding numbers in a finite chain of a SSH model with four or more bands are generally different if the system does not have a inversion symmetry. It would be interesting to see if this phenomenon also exists in 2D or 3D topological systems. Moreover unlike the edge states in the two-band SSH model, some of the edge states in the multi-band SSH models have non-vanishing energies. In fact, some of these edge states may lie very close to the band edge, see Figs. 5b and 5c in Appendix B for example. Hence, it is not certain whether they are still robust under perturbation. In particular, if we analyze the edge state of a right semi-infinite chain of a three-band SSH model, we find that there would be two edge states with energy $\Om =\pm t_0$ when $t_2 >t_1$ and there would be no edge state when $t_2 <t_1$. From our experience with the two-band SSH model, we would expect $t_2 = t_1$ to be a transition point and band crossing would occur in the energy spectrum of such a system.  However, it may be easily seen that this is not the case by choosing $(t_0, t_1, t_2 ) = (1, 2, 2)$.  As a result, we cannot be sure whether the system would indeed change from a trivial phase to a topological phase when we tune the value of $t_2$ from the $t_2 < t_1$ region to the $t_2 > t_1$ one.  One possible explanation is that a three-band SSH model does not have chiral symmetry and it can have topological phase only if the system has inversion symmetry.  At any rate, it is worthwhile to carry out a more detailed study to clarify this issue in the future.

\section*{Acknowledgments}
The work is supported in part by the Grants 109-2112-M-003-012 and 110B010029 of the Ministry of Science and Technology, Taiwan. The authors would like to thank Prof. Ming-Che Chang and Jhih-Shih You for helpful discussions.

\appendix
\section{Appendix A: Details of the calculation of the winding number in the four band SSH models}\label{App-A}
To calculate the winding number,  we must first find the wave functions of energy eigenstates:
\bea \label{SSH4 abcd 2}
&\;& \hskip -3.1cm \left|\Om_a(p) \right\ra ={\cal N}_a(p) \left\{ \ta_a(p), \tb_a(p), \tg_a(p), \td_a(p) \right\}^T,
\eea
with $a=1, \ldots, 4$. From eq.~(\ref{SSH4 abcd 2}), we see that they are given by
\bea
&\;& \hskip -3.1cm \ta_a(p) = \Bigl\{ t_3 \left( \Om_a^2(p) - t_1^2 \right) {\rm e}^{-ip} + t_0 t_1 t_2  \Bigr\}, \cr
&\;& \hskip -3.1cm \tb_a(p) = \Om_a(p) \Bigl\{  - t_1 t_2 - t_0 t_3 {\rm e}^{-ip}  \Bigr\} , \cr
&\;& \hskip -3.1cm \tg_a(p) = \Bigl\{ t_2 \left( \Om_a^2(p) - t_0^2 \right) + t_0 t_1 t_3 {\rm e}^{-ip}  \Bigr\}, \cr
&\;& \hskip -3.1cm \td_a(p) =  \Om_a(p)\Bigl\{ t_0^2 + t_1^2 - \Om_a^2(p) \Bigr\}.
\eea
Here, ${\cal N}_a(p)$ is a proper normalization constant that enforces the normalization condition $\left\la \Om_a(p)\, \right|\left. \Om_a(p) \right\ra = 1.$  In other words,
\bea
&\;& \hskip -3.1cm {\cal N}_a(p) = \left\{ \Bigl| \ta_a(p) \Bigr|^2 + \left|\tb_a(p) \right|^2 + \Bigl|\tg_a(p) \Bigr|^2 + \left|\td_a(p) \right|^2 \right\}^{-1/2}.
\eea
Moreover, we must choose $\td_a(p)$ to be real, which is equivalent to choosing a gauge. The reason for such a choice is that we should express $\ta_a(p), \tb_a(p)$ and $\tg_a(p)$ in terms of $\td_a(p)$, which in turn is coming from the BC on the left edge: $D_0 = 0$.  We will confirm that this is indeed the correct choice by checking the bulk edge correspondence both analytically and numerically later on.  By direct calculation, we see
\bea
&\;& \hskip -3.1cm  \left| \pd_p \Om(p) \right\ra ={\cal N}'_a(p) \left\{ \ta_a(p), \tb_a(p), \tg_a(p), \td_a(p) \right\}^T + {\cal N}_a(p) \left\{ \ta'_a(p), \tb'_a(p), \tg'_a(p), \td'_a(p) \right\}^T,
\ee
and hence
\bea
&\;& \hskip -3.1cm \left\la \Om_a(p) \right | \left. \pd_p \Om_a(p) \right\ra = \frac{ {\cal N}'_a(p)}{{\cal N}_a(p)}
+ {\cal N}_a(p)^2 \left\{ \ta^{*}_a(p) \ta'_a(p) + \tb^{*}_a(p) \tb'_a(p) + \tg^{*}_a(p) \tg'_a(p) + \td^{*}_a(p)  \td'_a(p) \right\}, \cr
&\;& \hskip -3.1cm \left\la \pd_p \Om_a(p) \right | \left. \Om_a(p) \right\ra = \frac{ {\cal N}'_a(p)}{{\cal N}_a(p)}
+ {\cal N}_a(p)^2 \left\{ \ta'^{*}_a(p) \ta_a(p) + \tb'^{*}_a(p) \tb_a(p) + \tg'^{*}_a(p) \tg_a(p) + \td'^{*}_a(p)  \td_a(p) \right\}.
\eea
As a result, we have
\bea
&\;& \hskip -3.1cm \left\la \Om_a(p) \right | \left. \pd_p \Om_a(p) \right\ra - \left\la \pd_p \Om_a(p) \right | \left. \Om_a(p) \right\ra =
\frac{2i {\rm Im} \left\{ \ta^{*}_a(p) \ta'_a(p) + \tb^{*}_a(p) \tb'_a(p) + \tg^{*}_a(p) \tg'_a(p) + \td^{*}_a(p)  \td'_a(p) \right\} }{ \Bigl| \ta_a(p) \Bigr|^2 + \left|\tb_a(p) \right|^2 + \Bigl|\tg_a(p) \Bigr|^2 + \left|\td_a(p) \right|^2}.
\eea
From eq.~(\ref{SSH4 Secular eq 2}), we may achieve the following identity:
\bea \label{Epp}
&\;& \hskip -3.1cm 2  \Om_a(p)  \Om'_a(p)\left\{ 2\Om_a^2(p) - \left(t_0^2 + t_1^2 + t_2^2 + t_3^2 \right) \right\} + 2 t_0 t_1 t_2 t_3  \sin p = 0.
\eea
Making use of all these relations, we will arrive at
\bea
&\;& \hskip -3.1cm \frac{i}{2} \Bigl\{ \left\la \Om_a(p) \right | \left. \pd_p \Om_a(p) \right\ra - \left\la \pd_p \Om_a(p) \right | \left. \Om_a(p) \right\ra  \Bigr \}  = \frac{ H(p, a) }{G(p, a) }.
\eea
Here,
\bea
&\;& \hskip -3.1cm G(p, a) = G_6(p) \Om_a^6(p) + G_4(p) \Om_a^4(p)+ G_2(p) \Om_a^2(p)+ G_0(p), \cr
&\;& \hskip -3.1cm H(p, a) = H_6(p) \Om_a^6(p) + H_4(p) \Om_a^4(p)+ H_2(p) \Om_a^2(p)+ H_0(p), \nn
\eea
with
\bea
&\;& \hskip -3.1cm G_6(p) = 1, \quad G_4(p) =  -2t_0^2 - 2 t_1^2 + t_2^2 + t_3^2, \cr
&\;& \hskip -3.1cm G_2(p) = - \left( t_0^2 + t_1^2 \right)^2 - 2 t_0^2 t_2^2 -2 t_1^2 t_3^2 + t_1^2 t_2^2 + t_0^2 t_3^2 + 6 t_0 t_1 t_2 t_3 \cos p, \cr
&\;& \hskip -3.1cm G_0(p) =  \left( t_0^2 + t_1^2 \right) \left( t_0^2 t_2^2 + t_1^2 t_3^2 - 2t_1 t_3 \cos p \right),
\eea
and
\bea
&\;& \hskip -3.1cm H_6(p) = -1, \quad H_4(p) =  2t_0^2 + 2 t_1^2 -t_2^2, \cr
&\;& \hskip -3.1cm H_2(p) =  - \left( t_0^2 + t_1^2 \right)^2 + 2 t_0^2 t_2^2 - t_1^2 t_2^2 - 3 t_0 t_1 t_2 t_3 \cos p, \cr
&\;& \hskip -3.1cm H_0(p) =  - t_0 t_2 \left( t_0^2 + t_1^2 \right) \left( t_0 t_2 - t_1 t_3 \cos p \right).
\eea
Note that the dependence on the band index $a$ only shows up through $\Om_a(p)$. It turns out that  a direct calculation would be too tedious to carry out even using mathematica.  Thus, it is important that we exploit this property to simplify the calculation of the winding number.

Since $\Om_a(p)$ satisfies eq.~(\ref{SSH4 Secular eq 2}), we may use the identity to simplify $G(p, a)$ and $H(p, a)$.  The best way to utilize it is to consider the polynomial division of $G(p, a)$ and $H(p, a)$ in $\Om_a(p)$ by
\bea
K(p, a) = \Om_a^4(p) - \left(t_0^2 + t_1^2 + t_2^2 + t_3^2 \right)\Om_a^2(p) + \left( t_0^2 t_2^2 + t_1^2 t_3^2 - 2 t_0 t_1 t_2 t_3  \cos p \right),
\eea
which is the polynomial of degree four in $\Om_a(p)$ that appears on the left hand side of eq.~(\ref{SSH4 Secular eq 2}).  We only need to keep the remainders of the polynomial division, and $G(p, a)$ and $H(p, a)$ will be reduced to $\Gt(p, a)$ and $\Ht(p, a)$, respectively.  It is straightforward to see that they are both polynomials of degree two in $\Om_a(p)$. The coefficients are
\bea
&\;& \hskip -3.1cm \Gt_2(p) = - 2 \left( t_0^2 t_2^2 - t_1^2 t_2^2 - t_2^4 - t_0^2 t_3^2 + t_1^2 t_3^2 - 2 t_2^2 t_3^2 - t_3^4 - 4 t_0 t_1 t_2 t_3 \cos p \right), \cr
&\;& \hskip -3.1cm \Gt_0(p) =  2\left( t_0^2 + t_1^2 - t_2^2 - t_3^2 \right) \left( t_0^2 t_2^2 + t_1^2 t_3^2 - 2t_1 t_3 \cos p \right).
\eea
and
\bea
&\;& \hskip -1.1cm \Ht_2(p) = 2 t_0^2 t_2^2 - 2 t_1^2 t_2^2 - 2 t_2^4 + t_1^2 t_3^2 - 3 t_2^2 t_3^2 - t_3^4 - 5 t_0 t_1 t_2 t_3 \cos p, \\
&\;& \hskip -1.1cm \Ht_0(p) = -2 t_0^4 t_2^2 - 2 _0^2 t_1^2 t_2^2 + 2 t_0^2 t_2^4 - t_0^2 t_1^2 t_3^2 - t_1^2 t_3^4 + t_0^2 t_2^2 t_3^2 + 2 t_1^2 t_2^2 t_3^2 + t_1^2 t_3^4 +  t_0 t_1 t_2 t_3 \left( 3 t_0^2 + 3 t_1^2 - 4 t_2^2 - 2 t_3^2 \right) \cos p. \nn
\eea
We may then express the integrand of the winding number by
\bea
&\;& \hskip -1.1cm \sum_{a=1,2} \frac{\Ht_2(p) \Om_a^2(p) + \Ht_0(p) } {\Gt_2(p) \Om_a^2(p) + \Gt_0(p) } \cr
&\;& \hskip -1.6cm  = \frac{2 \Gt_2(p) \Ht_2(p) \Bigl[\Om_1^2(p) \Om_2^2(p) \Bigr] + \Bigl[ \Gt_2(p)  \Ht_0(p) + \Gt_0(p)  \Ht_2(p) \Bigr]  \Bigl[ \Om_1^2(p) + \Om_2^2(p) \Bigr] + 2 \Gt_0(p) \Ht_0(p) } {\Gt_2^2(p) \Bigl[\Om_1^2(p) \Om_2^2(p) \Bigr] + \Gt_2(p) \Gt_0(p)\Bigl[\Om_1^2(p) + \Om_2^2(p) \Bigr] + \Gt_0^2(p) }.
\eea
From eq.~(\ref{SSH4 Secular eq 1}), it may be seen that $\Om_1^2(p)$ and $ \Om_2^2(p)$ satisfy the following relations:
\bea
&\;& \hskip -3.1cm \Om_1^2(p) + \Om_2^2(p) = t_0^2 + t_1^2 + t_2^2 + t_3^2; \cr
&\;& \hskip -3.1cm \Om_1^2(p) \Om_2^2(p) =  t_0^2 t_2^2 + t_1^2 t_3^2 - 2 t_0 t_1 t_2 t_3  \cos p.
\eea
Making use of these result, we may obtain the final expression for the winding number, which is given in eq.~(\ref{Winding-number-SSH4}).

\section{Appendix B: Six-band SSH models}\label{App-B}
We may follow similar procedures taken previously to study the six-band SSH model. Since the reasoning has been laid out explicitly in the previous section, we will only summarize the main results here. We first consider a right semi-infinite chain starting from the site $A_1$ so that we may investigate the edge states. The Hamiltonian is given by:
\bea
&\;& \hskip -3.1cm H_{\rm SSH-6}=\sum_{j=1}^{\infty}\left\{ t_0  {\bf A}_j^\dag {\bf B}_j  +t_1 {\bf B}_j^\dag {\bf C}_j + t_2 {\bf C}_j^\dag {\bf D}_j + t_3 {\bf D}_j ^\dag  {\bf E}_{j}  + t_4 {\bf E}_j^\dag {\bf F}_j + t_5 {\bf F}_j ^\dag  {\bf A}_{j+1}\right\} + \mbox{ h. c.}.
\eea
The energy eigenstates would satisfy the following EOM:
\bea \label{SSH6 EOM 1}
&\;& \hskip -3.1cm \Om A_j - \left(t_0 B_j + t_5 F_{j-1} \right) =0, \cr
&\;& \hskip -3.1cm \Om B_j - \left(t_0 A_j + t_1 C_j \right) =0, \cr
&\;& \hskip -3.1cm \Om C_j - \left(t_1 B_j + t_2 D_j \right) =0, \cr
&\;& \hskip -3.1cm \Om B_j - \left(t_2 D_j + t_3 E_j \right) =0, \cr
&\;& \hskip -3.1cm \Om C_j - \left(t_3 E_j + t_4 F_j \right) =0, \cr
&\;& \hskip -3.1cm \Om D_j - \left(t_4 F_j + t_0 A_{j+1}\right) =0,
\eea
which is again a linear recurrence relation with constant coefficients.  After simplification, the OBC is given by
\bea \label{SSH6 Simplified BC 1}
&\;& \hskip -3.1cm F_0 =0.
\eea
To find the solutions, we let
\bea
&\;& \hskip -3.1cm A_j =\a s^j, B_j=\b s^j, C_j =\g s^j, D_j=\d s^j, E_j =\eps s^j, F_j=\et s^j.
\eea
As a result, we have
\bea \label{abcdef 1}
&\;& \hskip -3.1cm \Om \a - \left( t_0 \b + t_5 s^{-1} \et \right) = 0, \cr
&\;& \hskip -3.1cm \Om \b - \left( t_0 \a + t_1 \g \right) = 0, \cr
&\;& \hskip -3.1cm \Om \g - \left( t_1 \b + t_2 \d \right) = 0, \cr
&\;& \hskip -3.1cm \Om \d - \left( t_2 \g + t_3 \eps \right)= 0 \cr
&\;& \hskip -3.1cm \Om \eps - \left( t_3 \d + t_4 \et \right) = 0, \cr
&\;& \hskip -3.1cm \Om \et - \left( t_4 \eps + t_5 s \a \right)= 0.
\eea
We may obtain from the above equations the following secular equation:

\bea \label{SSH6 Secular eq 1}
&\;& \hskip -1.1cm \Om^6 - \left(t_0^2 + t_1^2 + t_2^2 + t_3^2 + t_4^2 + t_5^2 \right)\Om^4
+ \left(t_0^2 t_2^2 + t_0^2 t_3^2 +t_0^2 t_4^2 + t_1^2 t_3^2 + t_1^2 t_4^2 + t_1^2 t_5^2 + t_2^2 t_4^2 + t_2^2 t_5^2 + t_3^2 t_5^2 \right)\Om^2  \cr
&\;& \hskip -1.1cm - \left( t_0^2 t_2^2 t_4^2 + t_1^2 t_3^2 t_5^2 + 2 t_0 t_1 t_2 t_3 t_4 t_5 u \right) = 0.
\eea
Note that the equation is sextic in $\Om$, so there are six roots. Similar to the four-band case, the most general forms of $A_j, B_j, C_j, D_j, E_j$ and $F_j$ are given by
\bea \label{SSH6 general solution}
&\;& \hskip -3.1cm A_j = \a_+ s^j + \a_- s^{-j}, B_j = \b_+ s^j + \b_- s^{-j},  C_j = \g_+ s^j + \g_- s^{-j},   \cr
&\;& \hskip -3.1cm D_j = \d_+ s^j + \d_- s^{-j}, E_j = \eps_+ s^j + \eps_- s^{-j},  F_j = \et_+ s^j + \et_- s^{-j}.
\eea
If we focus on the edge states and assume $|s|<1$, then physical solutions exist only if $\a_- = \b_- = \g_- = \d_- = \eps_- = \et_- =0$. After some algebra, we find the solutions are given by
\bea \label{SSH6 Edge state solution}
&\;& \hskip -3.1cm s = s_0,\;  \Om = 0, \cr
&\;& \hskip -3.1cm s = s^*_{+},\;  \Om =\pm \left\{ \frac{ t_0^2+t_1^2+t_2^2+t_3^2 -\sqrt{(t_0^2+t_1^2+t_2^2+t_3^2)^2-4(t_1^2 t_3^2 + t_0^2 t_2^2+  t_0^2 t_3^2 )} }{2} \right\}^{1/2}, \cr
&\;& \hskip -3.1cm s = s^*_{-},\;  \Om =\pm \left\{ \frac{ t_0^2+t_1^2+t_2^2+t_3^2 +\sqrt{(t_0^2+t_1^2+t_2^2+t_3^2)^2-4(t_1^2 t_3^2 + t_0^2 t_2^2+  t_0^2 t_3^2 )} }{2} \right\}^{1/2}.
\eea
Here,
\be
s_0 = -\frac{t_0 t_2 t_4}{t_1 t_3 t_5},
\;s^*_{\pm} = \frac{t_3 t_4 \left\{t_0^2+t_1^2-t_2^2-t_3^2 \pm \sqrt{ (t_0^2+t_1^2+t_2^2+t_3^2 )^2 - 4(t_1^2 t_3^2 + t_0^2 t_2^2+  t_0^2 t_3^2 )  } \right\} }{2t_0 t_1 t_2 t_5}. \nn
\ee
From the above results, we see there are indeed six phases as expected.
\bee[label=\arabic*)]
\item $\left| s_0 \right|<1, |s^*_{+}|<1, |s^*_{-}|<1$: five edge states, one of which has zero energy.

\item $\left| s_0 \right|>1, |s^*_{+}|<1, |s^*_{-}|<1$: four edge states.

\item $\left| s_0 \right|<1$, either $|s^*_{+}|>1, |s^*_{-}|<1$ or $|s^*_{+}|<1, |s^*_{-}|>1$: three edge states, one of which has zero energy.

\item $\left| s_0 \right|>1$, either $|s^*_{+}|>1, |s^*_{-}|<1$ or $|s^*_{+}|<1, |s^*_{-}|>1$: two edge states.

\item $\left| s_0 \right|<1, |s^*_{+}|>1, |s^*_{-}|>1$: one zero energy edge state.

\item $\left| s_0 \right|>1, |s^*_{+}|>1, |s^*_{-}|>1$: no edge states.
\eee

The Bloch Hamiltonian here is given by
\be\label{SSH-6}
\HH_{\rm SSH-6}=\left(
\begin{matrix}
0 & t_0 & 0 & 0 & 0 & t_5 {\rm e}^{-ip} \cr
t_0 & 0 & t_1 & 0 & 0 & 0 \cr
0 & t_1 & 0 & t_2 & 0 & 0  \cr
0 & 0 & t_2 & 0 & t_3 & 0  \cr
0 & 0 & 0 & t_3 & 0 & t_4  \cr
t_5 {\rm e}^{ip} & 0 & 0 & 0 & t_4 & 0 \cr
\end{matrix}
\right).
\ee
From eqs.~(\ref{SSH6 Secular eq 1}), we see the secular equation becomes
\bea \label{SSH6 Secular eq 3}
&\;& \hskip -1.1cm \Om^6 - \left(t_0^2 + t_1^2 + t_2^2 + t_3^2 + t_4^2 + t_5^2 \right)\Om^4
+ \left(t_0^2 t_2^2 + t_0^2 t_3^2 + t_1^2 t_3^2 + t_0^2 t_4^2 + t_1^2 t_4^2 + t_2^2 t_4^2 + t_1^2 t_5^2 + t_2^2 t_5^2 + t_3^2 t_5^2 \right)\Om^2  \cr
&\;& \hskip -1.1cm - \left( t_0^2 t_2^2 t_4^2 + t_1^2 t_3^2 t_5^2 + 2 t_0 t_1 t_2 t_3 t_4 t_5 \cos p \right) = 0.
\eea
Here, the chiral operator is given by
\be\label{Pi}
\Pi=\left(
\begin{matrix}
1 & 0 & 0 & 0 & 0 & 0 \cr
0 & -1 & 0 & 0 & 0 & 0 \cr
0 & 0 & 1 & 0 & 0 & 0 \cr
0 & 0 & 0 & -1 & 0 & 0 \cr
0 & 0 & 0 & 0 & 1 & 0 \cr
0 & 0 & 0 & 0 & 0 & -1 \cr
\end{matrix}
\right).
\ee
Hence, the system also belongs to the BDI class and non-zero energy states always appear in pairs.  The six solutions to eq.~(\ref{SSH6 Secular eq 3}) are given by $\Om_1(p) = -\om_3(p), \Om_2(p) = -\om_2(p), \Om_3(p) = -\om_1(p), \Om_4(p) = \om_1(p), \Om_5(p) = \om_2(p)$, and $\Om_6(p) = \om_3(p)$.

The corresponding wave functions are given by
\bea \label{SSH6 abcdef 3}
&\;& \hskip -3.1cm \left|\Om_a(p) \right\ra ={\cal N}_a(p) \left\{ \ta_a(p), \tb_a(p), \tg_a(p), \td_a(p), \tilde{\eps}_a(p), \tilde{\et}_a(p) \right\}^T,
\eea
with $a=1, \ldots, 6$. Here,
\bea
&\;& \hskip -3.1cm \ta_a(p) = {\rm e}^{-ip} \Bigl\{  t_1 t_3 \left(t_0 t_2 t_4 {\rm e}^{ip} + t_1 t_3 t_5 \right) -  t_5\left(t_1^2 + t_2^2 +  t_3^2 \right) \Om^2_a(p) + t_5 \Om^4_a(p)  \Bigr\} , \cr
&\;& \hskip -3.1cm \tb_a(p) = -\Om_a(p){\rm e}^{-ip} \Bigl\{  t_1 t_2 t_3 t_4 {\rm e}^{ip} - t_0 t_5\left(t_2^2 + t_3^2 \right)  + t_0 t_5 \Om^2_a(p)  \Bigr\} , \cr
&\;& \hskip -3.1cm \tg_a(p) = \Bigl\{ t_0 t_3 \left(-t_0 t_2 t_4 - t_1 t_3 t_5 {\rm e}^{-ip} \right) + \left(t_2 t_3 t_4 + t_0 t_1 t_5 {\rm e}^{-ip} \right) \Om^2_a(p)  \Bigr\}, \cr
&\;& \hskip -3.1cm \td_a(p) = -\Om_a(p)\Bigl\{-t_0^2 t_3 t_4 - t_1^2 t_3 t_4 + t_0 t_1 t_2 t_5 {\rm e}^{-ip} + t_3 t_4 \Om^2_a(p)  \Bigr\}, \cr
&\;& \hskip -3.1cm \tilde{\eps}_a(p) = \Bigl\{ t_0 t_2 \left(t_0 t_2 t_4 + t_1 t_3 t_5 {\rm e}^{-ip} \right) - t_4\left(t_0^2 + t_1^2 + t_2^2 \right)  \Om^2_a(p) + t_4 \Om^4_a(p)  \Bigr\}, \cr
&\;& \hskip -3.1cm \tilde{\et}_a(p) = -\Om_a(p)\Bigl\{ t_1^2 t_3^2 + t_0^2 \left(t_2^2 + t_3^2 \right) - \left(t_0^2 + t_1^2 + t_2^2 + t_3^2 \right) \Om^2_a(p) + \Om^4_a(p) \Bigr\}.
\eea
Again, we must choose $\tilde{\et}_a(p)$ to be real.

The rest of the procedures are more or less the same as those in the four-band case, except that the calculation is much more tedious.  Make use of eq.~(\ref{SSH6 Secular eq 3}) to simplify the expression and the integrand of the winding number may be expressed as:
\bea
&\;& \hskip -1.1cm \sum_{a=1,2,3} \frac{\Ht_4(p) \Om_a^4(p) +\Ht_2(p) \Om_a^2(p) + \Ht_0(p) } {\Gt_4(p) \Om_a^4(p) + \Gt_2(p) \Om_a^2(p) + \Gt_0(p) }.
\eea

Eventually, the winding number is given by
\bea
&\;& \hskip -3.1cm \n = \frac{1}{4 \pi} \int_{-\pi}^{\pi} dp\, \Biggl\{ 4 + \g_1+ \bg_1 + \g_2 + \bg_2 \Biggr\}.
\eea
Here, $\g_1 = \frac{ t_1 t_3 t_5 }{ t_1 t_3 t_5 + t_0 t_2 t_4 {\rm e}^{ip} }$, $\g_2  = \frac{2 t_1 t_2 \left( t_0^2 t_5^2 {\rm e}^{ip} + t_3^2 t_4^2  {\rm e}^{-ip} \right)}{ t_1 t_2 t_0^2 t_5^2 {\rm e}^{ip} - t_0 t_3  t_4 t_5 \left(t_0^2 + t_1^2 - t_2^2 - t_3^2 \right) - t_1 t_2 t_3^2 t_4^2   {\rm e}^{-ip} }$ and $\bg_1, \bg_2$ are their complex conjugates, respectively.  From the expressions for $\g_1$ and $\g_2$, we may find the following analytic form for the winding number:
\bea\label{SSH6-winding-number}
&\;& \hskip -1.1cm \n = 2 + \th\left( \left|t_1 t_3 t_5 \right| - \left|t_0 t_2 t_4 \right| \right) - \left[ 2\th \left( t_3^2 t_4^2 - t_0^2 t_5^2 \right) -1 \right]
\th\left[ \left| t_1 t_2 \left( t_3^2 t_4^2 - t_0^2 t_5^2 \right) \right| - \left| t_0 t_3 t_4 t_5 \left( t_0^2 + t_1^2 - t_2^2 - t_3^2 \right) \right| \right].
\eea
One can easily check that the above result is equivalent to that obtained by analyzing the edge states.

To carry out a numerical check of the above predictions, we must again consider a finite chain. Similar to the semi-infinite chain, the site by the left boundary is chosen to be $A_1$. On the other hand, the site by the right boundary depends on the total number of sites $N_{\rm tot}$ in the system. Consequently, the conditions for the number of edge states on the right boundary would change accordingly.
\bee[label=\arabic*)]
\item  $N_{\rm tot} = 6N,$ the right BC: $A_{N+1}=0$. $\tt_0 = t_4, \tt_1 = t_3, \tt_2 = t_2, \tt_3 = t_1, \tt_4 = t_0$ and $\tt_5 = t_5$. \hfill\break
The corresponding characteristic equation is
\bea
&\;& \hskip -3.1cm   t_0 t_1 t_2 t_3 t_4 U_{N}(u) + t_5 \left[ \Om^4 - \left( t_1^2 + t_2^2 + t_3^2 \right) \Om^2 + t_1^2  t_3^2 \right] U_{N-1}(u) = 0,
\eea
Similar to the four-band case, unless $t_0 = t_4$ and $t_1 = t_3$  so that each unit cell of the system has a space inversion symmetry with respect to the center position of the cell, the numbers of edge states on the left and right boundaries would generally be different.

\item $N_{\rm tot} = 6N+1$, the right BC: $B_{N+1}=0$. $\tt_0 = t_5, \tt_1 = t_4, \tt_2 = t_3, \tt_3 = t_2, \tt_4 = t_1$ and $\tt_5 = t_0$.\hfill\break
The corresponding characteristic equation is
\bea
&\;& \hskip -3.1cm   t_1 t_2 t_3 t_4 U_{N}(u) + t_0 t_5 \left[ \Om^2 - \left( t_2^2 + t_3^2 \right)  \right] U_{N-1}(u) = 0,
\eea

\item $N_{\rm tot} = 6N+2$, the right BC: $C_{N+1}=0$. $\tt_0 = t_0, \tt_1 = t_5, \tt_2 = t_4, \tt_3 = t_3, \tt_4 = t_2$ and $\tt_5 = t_1$.\hfill\break
The corresponding characteristic equation is
\bea
&\;& \hskip -3.1cm    t_2 t_3 t_4  \left[ \Om^2 - t_0^2  \right]  U_{N}(u) + t_0 t_1 t_5 \left[ \Om^2 -  t_3^2 \right] U_{N-1}(u) = 0,
\eea

\item $N_{\rm tot} = 6N+3$, the right BC: $D_{N+1}=0$. $\tt_0 = t_1, \tt_1 = t_0, \tt_2 = t_5, \tt_3 = t_4, \tt_4 = t_3$ and $\tt_5 = t_2$.\hfill\break
The corresponding characteristic equation is
\bea
&\;& \hskip -3.1cm   t_0 t_1 t_2 t_5 U_{N-1}(u) + t_3 t_4 \left[ \Om^2 - \left( t_0^2 + t_1^2 \right)  \right] U_{N}(u) = 0,
\eea

\item $N_{\rm tot} = 6N+4$, the right BC: $E_{N+1}=0$. $\tt_0 = t_2, \tt_1 = t_1, \tt_2 = t_0, \tt_3 = t_5, \tt_4 = t_4$ and $\tt_5 = t_3$.\hfill\break
The corresponding characteristic equation is
\bea
&\;& \hskip -3.1cm   t_0 t_1 t_2 t_3 t_5 U_{N-1}(u) + t_5 \left[ \Om^4 - \left( t_0^2 + t_1^2 + t_2^2  \right) \Om^2 + t_1^2  t_3^2 \right] U_{N}(u) = 0,
\eea

\item $N_{\rm tot} = 6N-1$, the right BC: $F_{N}=0$. $\tt_0 = t_3, \tt_1 = t_2, \tt_2 = t_1, \tt_3 = t_0, \tt_4 = t_5$ and $\tt_5 = t_4$.\hfill\break
The corresponding characteristic equation is
\bea
&\;& \hskip -3.1cm   s^N - s^{-N} = 0,
\eea

\eee

\begin{figure}[hbt!]
\centering
\includegraphics[width=0.60\textwidth]{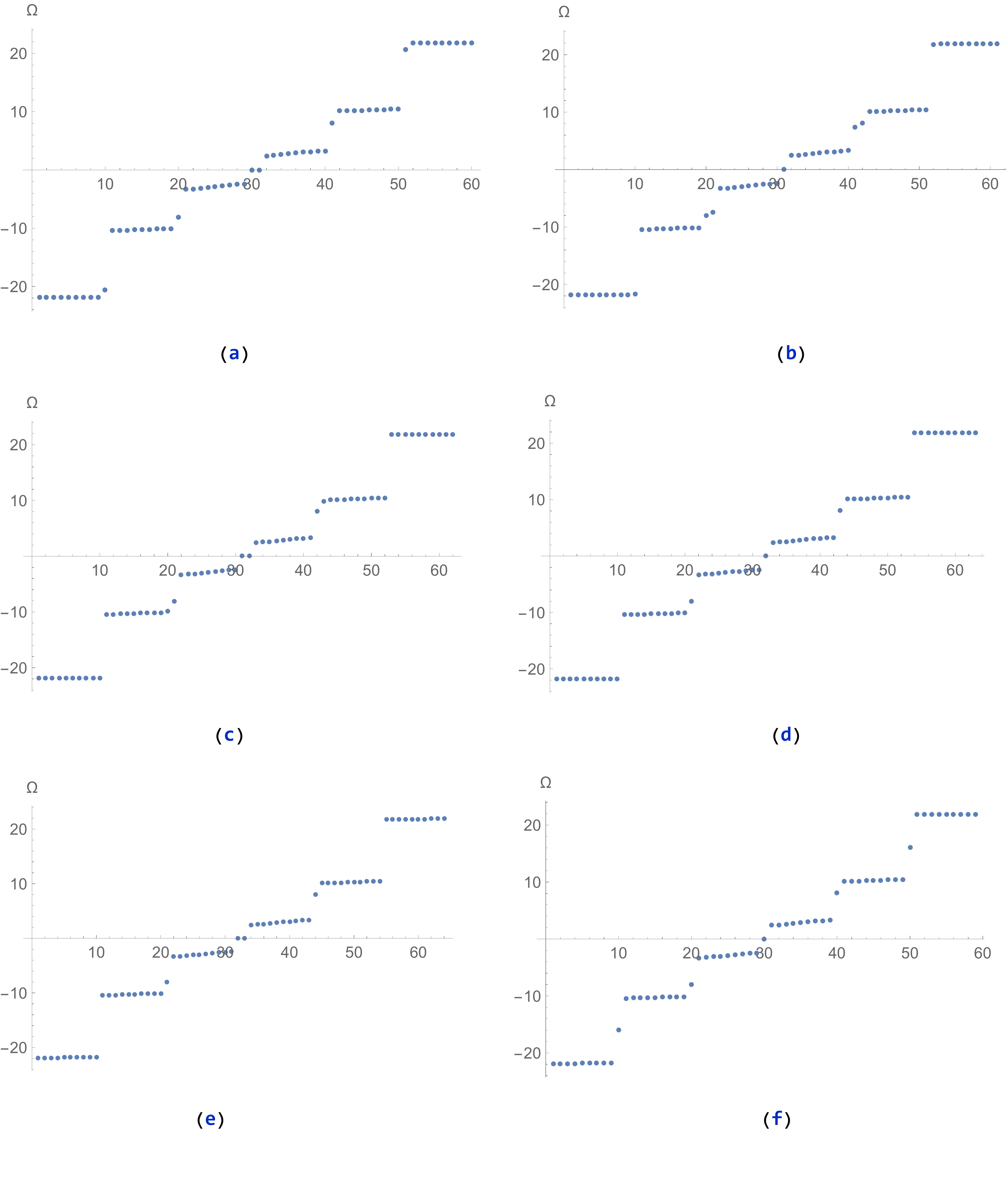}
\caption{The parameters are chosen to be $\left( t_0, t_1, t_2, t_3, t_4,t_5\right) = (7, 4, 1, 16, 13, 10)$ so that $\n_{\rm L} = 3$. The energy spectra for the four cases that $N_{\rm tot} = 60, 61, 62, 63, 64$ and $59$ are shown. (a)$N_{\rm tot} =60$, $\n_{\rm R} = 3$. There are six edge states in total, two of which have approximately zero energy.  (b) $N_{\rm tot} = 61$, $\n_{\rm R} = 4$. There are seven edge states in total, one of which have approximately zero energy. , (c) $N_{\rm tot} = 62$, $\n_{\rm R} = 3$. There are six edge states in total, two of which have approximately zero energy.  (d)$N_{\rm tot} = 63$, $\n_{\rm R} = 0$. There are three edge states in total, one of which have approximately zero energy. (e) $N_{\rm tot} = 64$, $\n_{\rm R} = 1$. There are four edge states in total, two of which have approximately zero energy.  (f)$N_{\rm tot} = 59$, $\n_{\rm R} = 2$. There are five edge states in total, one of which have approximately zero energy. }  \label{fig5 SSH6}
\end{figure}

\begin{figure}[bt!]
\centering
\includegraphics[width=0.60\textwidth]{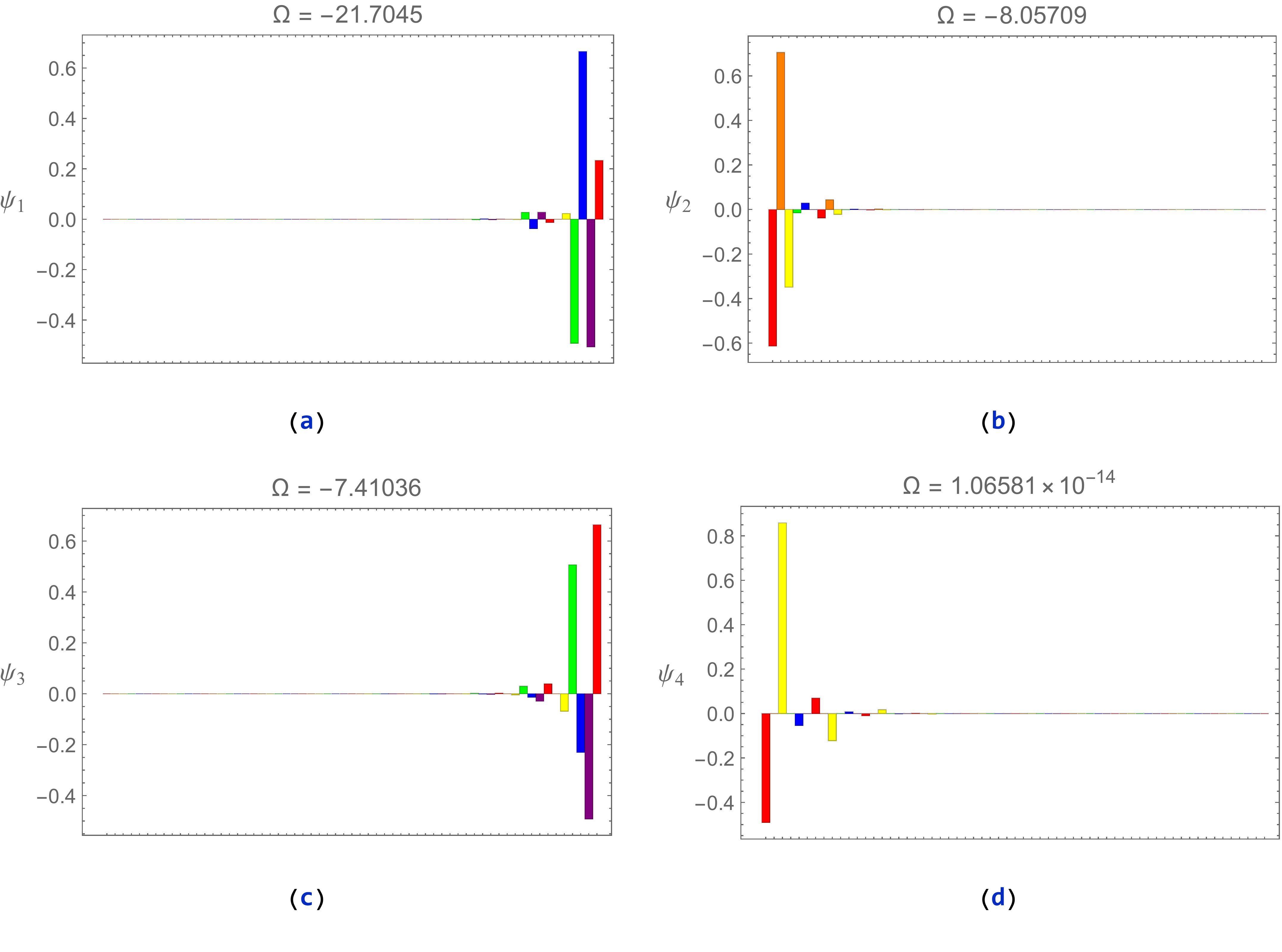}
\caption{Here we show the wave functions for the edge states in the case that $\left( t_0, t_1, t_2, t_3, t_4,t_5\right) = (7, 4, 1, 16, 13, 10)$ with $N_{\rm tot} = 61$.(a) The wave functions for the edge state with the lowest energy is denoted $\psi_1$. It is a right edge state.  (b) The wave functions for the edge state with the second lowest energy is denoted $\psi_2$. It is a left edge state. (c) The wave functions for the edge state with the third lowest energy is denoted $\psi_3$. It is a right edge state. (d) The wave functions of the zero energy edge state is denoted $\psi_4$. It is also a left edge state}  \label{fig6 SSH6}
\end{figure}

In Fig.~\ref{fig5 SSH6}, we show the energy spectra for a six-band SSH model with the parameters $\left( t_0, t_1, t_2, t_3, t_4,t_5\right) = (7, 4, 1, 16, 13, 10)$ so that $\n_{\rm L} = 3$ with $N_{\rm tot} = 60, 61, 62, 63, 64, 59$ respectively.  The corresponding $\n_{\rm R} = 3, 4, 3, 0, 1, 2$. They are all calculated from eq.~(\ref{SSH6-winding-number}). Except for the case that $N_{\rm tot} = 61, 62$, it may be seen from the figures that $\n_{\rm L} +\n_{\rm R}$ correctly predicts the number of edge states. In Fig.~\ref{fig6 SSH6}, we show the wave functions of the four lowest energy edge states. Because of the chiral symmetry, we know there are three more edge states with positive energy. One of them is a left edge state and the other two right edge states.  Thus, there are indeed three left and four right edge states in this case.  The energies of the edge states are consistent with the prediction of eq.~(\ref{SSH6 Edge state solution}) as well. Therefore, we again confirm the prescription we proposed is correct by checking the bulk edge correspondence.

\begin{thebibliography}{99}
\bibitem{Review} M. Z. Hasan and C. L. Kane, "Topological insulators," Rev. Mod. Phy. {\bf 82}, (2010); X.-L. Qi and S.-C. Zhang, "Topological insulators and superconductors," Rev. Mod. Phy. {\bf 83}, (2011);	J. Alicea, Rep. Prog. Phys. {\bf 75}, 076501 (2012).

\bibitem{SSH} W. P. Su, J. R. Schrieffer, and A. J. Heeger, "Solitons in Polyacetylene," Phys. Rev. Lett. {\bf 42}, 1698, (1979).

\bibitem{Asboth} J. K. Asboth, L. Oroszlany, and A. Palyi, "A Short Course on Topological Insulators," arXiv:1509.02295.

\bibitem{Trimer} W. P. Su and J. R. Schrieffer, ``Fractionally Charged Excitations in Charge-Density-with Commensurability 3,'' Phys. Rev. Lett. {\bf 46}, 738 (1981); W. P. Su, ''Fractionally charged kinks in a I:3 Peierls system,''
Phys. Rev. B {\bf 27}, 370 (1983).

\bibitem{Three-band} V. M. Martinez Alvarez and M. D. Coutinho-Filho Phys. Rev. A {\bf 99}, 013833 (2019); Y. Zhang, B. Ren, Y. Li, and F. Ye, Optics Express {\bf 29}, 42827, (2021).

\bibitem{PA-ladder} W. Song,  H. Han,  J. Wu  and  M. Xie, ``Ladder-like polyacetylene with excellent optoelectronic properties and regular architecture,'' Chem. Comm. {\bf 50}, 12899 (2014).


\bibitem{Zak} J. Zak, "Berry’s phase for energy bands in solids," Phys. Rev. Lett., {\bf 62}, 2747, (1989).

\bibitem{Berry} M. V. Berry,  "Quantal phase factors accompanying adiabatic changes," Proc. R. Soc. Lond. A. {\bf 392}, 45,  (1984).

\bibitem{Atala} M. Atala {\it et al.}, ``Direct Measurement of the Zak phase in Topological Bloch Bands,'' Nature Phys {\bf 9}, 795 (2013) .

\bibitem{Rice-Mele} Rice, M. J., and E. J. Mele, ``Elementary Excitations of a Linearly Conjugated Diatomic Polymer,'' Phys. Rev. Lett. {\bf 49}, 1455 (1982).

\bibitem{Rhim} J.-W. Rhim, J. Behrends, and J. H. Bardarson, ``Bulk-boundary correspondence from the intercellular Zak phase,'' Phys. Rev. B {\bf 95}, 035421 (2017).

\bibitem{Thouless} D.J. Thouless, "Quantization of particle transport," Phys. Rev. B {\bf 27}, 6083–6087 (1983)

\bibitem{polarization} R. D. King-Smith and D. Vanderbilt, "Theory of polarization of crystalline solids," Phys. Rev. B {\bf 47}, 1651(R) (1993); D. Vanderbilt and R. D. King-Smith, "Electric polarization as a bulk quantity and its relation to surface charge," Phys. Rev. B {\bf 48}, 4442 (1993).

\bibitem{Ext-SSH}  Han-Ting Chen, ,Chia-Hsun Chang and Hsien-chung Kao, `Connection between the winding number and the Chern number,'' Chinese Journal of Physics, {\bf 72}, 50-68, (2021).

\bibitem{Kudin}  K. N. Kudin, R. Car, and R. Resta, J. Chem. Phys. {\bf 126}, 234101 (2007).

\bibitem{Holonomy} R. Leone, J. Phys. A: Math. Theor. {\bf 44}, 295301 (2011).

\bibitem{W-loop} F. Wilczek and A. Zee, Phys. Rev. Lett. {\bf 52}, 2111 (1984).

\bibitem{H-application}  A. Ekert, M. Ericsson, P. Hayden, H. Inamori, J. A. Jones, D. K. L. Oi, and V. Vedral, J. Mod. Opt. {\bf 47}, 2501 (2000);  A. Recati, T. Calarco, P. Zanardi, J. I. Cirac, and P. Zoller, Phys. Rev. A {\bf 66}, 032309 (2002).

\bibitem{Periodic-table}  A. P. Schnyder, S. Ryu, A. Furusaki, and A. W. W. Ludwig, Phys. Rev. B {\bf 78}, 195125 (2008); A. Kitaev, AIP Conf. Proc. 1134, 22 (2009); A. P. Schnyder, S. Ryu, A. Furusaki, and A. W. W. Ludwig, AIP Conf. Proc. 1134, 10 (2009).

\bibitem{TKNN} D. J. Thouless, M.  Kohmoto, M. P. Nightingale,  and M. den Nijs, Phys. Rev. Lett. {\bf 49},  405 (1982);Y. Hatsugai, Phys. Rev. Lett. {\bf 71}, 3697 (1993).

\bibitem{Space-inversion} L. Fu and C. L. Kane, Phys. Rev. B 76, 045302 (2007).

\bibitem{TC-insulators}  L. Fu, Phys. Rev. Lett. {\bf 106}, 106802 (2011); T. Neupert and F. Schindler, ``Topological Crystalline Insulators,'' arXiv:1810.03484.


\end{thebibliography}
\end{document}